
\FRONTPAGE
\line{\hfill BROWN-HET-945}
\line{\hfill June 1994}
\vskip1.0truein
\titlestyle{{{\bf COSMIC STRINGS AND ELECTROWEAK SYMMETRY RESTORATION IN THE
TWO-HIGGS DOUBLET MODEL}}
\foot{Work supported in part by the Department of Energy under
contract DE-FG02-91ER40688-Task A}}
\bigskip
\author {Mark Trodden\foot{e-mail address: mtrodden@het.brown.edu}}
\vskip .25in
\item { }{\it Physics Department, Brown University, Providence, RI
02912, USA}
\bigskip
\abstract

We investigate the region of restored electroweak symmetry around a
superconducting cosmic string coupled to the Weinberg-Salam model with an extra
Higgs doublet. We show that the presence of the extra Higgs fields, and in
particular their CP violating relative phase, does not qualitatively affect the
results of Perkins and Davis[1] obtained for the standard electroweak theory.
This result is neccessary to ensure the validity of some recently proposed
string-mediated electroweak baryogenesis scenarios.

\endpage
{\bf \chapter{Introduction}}
It has recently been realised that the electroweak symmetry is restored in the
presence of a superconducting Grand Unified (GUT) cosmic string[1,2]. In
particular, for sufficiently large string currents, the unbroken phase of the
Weinberg-Salam theory exists in a region of macroscopic radius ($\sim
10^{-5}$m) around the central core of the defect. In such a region of restored
symmetry we expect that baryon number violating processes, due to the chiral
anomaly structure of the standard model, will be unsuppressed[3] and that their
rate will be comparable with that at the time of the electroweak phase
transition[4]. This presents the possibility that cosmic strings may play an
important role in the generation of the baryon asymmetry of the universe
(BAU)[5].

A specific scenario implementing the above ideas was presented in Ref.~5. This
model is particularly attractive for two principal reasons. Firstly, it does
not rely on the electroweak phase transition being first order, as do all other
current models for electroweak baryogenesis (EWBG). Secondly, there is a
possibility that this scenario may be viable using GUT strings, thus supporting
the existence of the same cosmic strings that may be responsible for the
formation of large scale structure in the universe[6-9]. However, as in all
other models of EWBG, this scenario requires a new source of CP violation in
addition to those found in the standard model[10-14]. The canonical way of
achieving this is to assume an extended Higgs sector - two Higgs doublets -
with a CP violating relative phase between them. Thus, it is the object of this
paper to investigate the restoration of the electroweak symmetry in such an
extension of the standard model and confirm that the predictions of Ref.~1,
neccessary for some imple

mentations of Ref.~5, are unaffected by the new Higgs physics.

The paper is organized as follows. In section~2 we present the model we shall
analyze in detail and derive the equations of motion. This model is an
extension of the model presented in[1] with a superconducting cosmic string
coupled to the two Higgs doublet electroweak theory. In section~3 we make a
specific ansatz for the field profiles and, in a certain approximation, derive
the radius of the region of restored electroweak symmetry, showing no
qualitative difference from the results of Ref.~1. Section~4 is devoted to a
numerical study of the one-Higgs case in order to justify the approximations
made in the analytical study. Finally, in section~5 we conclude and comment on
the relevance of our results for EWBG.

{\bf \chapter{A Specific Model}}
We shall use a simple extension of the model proposed in Ref.~1. Consider an
$SU(2)\times U(1)\times U(1)$ invariant gauge field theory. We identify $SU(2)$
with weak isospin, the first $U(1)$ with hypercharge and the final $U(1)$ with
a spontaneously broken symmetry giving rise to cosmic strings. Throughout the
analysis we neglect the fermionic sector of the theory and introduce an
additional scalar field which couples to hypercharge such that the string
becomes superconducting. The Higgs sector associated with the electroweak part
of the model is assumed to consist of two doublets with the general two-Higgs
potential.

Thus, the full Lagrangian for the model is
$$\eqalign{
{\cal L} = &({\tilde D}_{\mu}\Sigma)^{\dagger}({\tilde D}^{\mu}\Sigma) -
           {1\over 4}R_{\mu\nu}R^{\mu\nu} -{\lambda\over
4}(|\Sigma|^2-\eta^2)^2
          +({\hat D}_{\mu}\chi)^{\dagger}({\hat D}^{\mu}\chi) - m^2\chi^2 \cr
           &-{1\over 4}|\chi|^4
           +f|\Sigma|^2|\chi|^2
          +(D_{\mu}\Phi^i)^{\dagger}(D^{\mu}\Phi^i)
          -{1\over 4}F_{\mu\nu}F^{\mu\nu} -{1\over 4}W_{\mu\nu}^aW^{a\mu\nu}
          -V(\Phi_1,\Phi_2) \cr} \eqno\eq
$$
Here, $B_{\mu}$ is the hypercharge field, $W^a_{\mu}$ are the weak isospin
fields and $R_{\mu}$ is the new $U(1)$ field. The field strengths are
$R_{\mu\nu}=R_{[\nu,\mu]}$, $F_{\mu\nu}=B_{[\nu,\mu]}$,
$W^a_{\mu\nu}=W^a_{[\nu,\mu]} + g\varepsilon^{abc}W^b_{\mu}W^c_{\nu}$, the
covariant derivatives are given by ${\tilde D}_{\mu}=\partial_{\mu}-iqR_{\mu}$,
${\hat D}_{\mu}=\partial_{\mu}-{i\over 2}{\tilde g}'B_{\mu}$,
$D_{\mu}=\partial_{\mu}-{i\over 2}g{\bf \tau}.{\bf W}_{\mu}-{i\over 2}{\tilde
g}'B_{\mu}$, the index $i=1,2$ and the general two-Higgs potential is[15]
$$\eqalign{
V(\Phi_1,\Phi_2) = &\lambda_1(\Phi_1^{\dagger}\Phi_1 - {\nu_1^2 \over 2})^2 +
                   \lambda_2(\Phi_2^{\dagger}\Phi_2 - {\nu_2^2 \over 2})^2 +
\cr
                   &\lambda_3[(\Phi_1^{\dagger}\Phi_1 - {\nu_1^2 \over 2}) +
(\Phi_2^{\dagger}\Phi_2 - {\nu_2^2 \over 2})]^2 +
                   \lambda_4[(\Phi_1^{\dagger}\Phi_1)(\Phi_2^{\dagger}\Phi_2) -
(\Phi_1^{\dagger}\Phi_2)(\Phi_2^{\dagger}\Phi_1)] + \cr
                   &\lambda_5[{\rm Re}(\Phi_1^{\dagger}\Phi_2) -{\nu_1\nu_2
\over 2}\cos\xi]^2 +\lambda_6[{\rm Im}(\Phi_1^{\dagger}\Phi_2) -{\nu_1\nu_2
\over 2}\sin\xi]^2 \cr}\eqno\eq
$$
Thus, it is the scalar field, $\Sigma$, which forms the cosmic string and we
assume that the parameters of the theory are such that the field $\chi$
condenses in the string core leading to a superconducting current along the
string.

Now, following Witten[16], we assume that $\Sigma$ has the profile of a
Nielsen-Olesen[17] cosmic string and that the superconducting condensate obeys
$$
\chi = \chi_0(r,\theta)e^{i{\tilde \nu}(z,t)} \eqno\eq
$$
where we have adopted cylindrical polar coordinates $(r,\theta,z)$ with the
z-axis aligned with the string core. Here $\chi_0$ is the spatially-dependent
amplitude of the condensate and ${\tilde \nu}$ is the phase. It is ${\tilde
\nu}$ that acts as a source for the hypercharge field and since the condensate
is charged this results in a hypercharge current along the string. With this
assumption we may derive the equations of motion for the electroweak and Higgs
fields.
For the hypercharge field we have
$$
\partial_{\nu}B^{[\nu,\mu]} = 2\left[(D^{\mu}\Phi^i)^{\dagger}(-{i\over
2}g'\Phi^i) + \chi_0^{\dagger}{i\over 2}{\tilde g}'(i\partial^{\mu}{\tilde \nu}
-{i\over 2}{\tilde g}'B^{\mu})\chi_0 + ({\rm h.c.})\right] \eqno\eq
$$
For the $SU(2)$ gauge fields
$$\eqalign{
\partial_{\mu}&(W^{a[\nu,\mu]}+g\varepsilon^{ade}W^{d\mu}W^{e\nu}) =  \cr
&2\left[(D^{\mu}\Phi^i)^{\dagger}(-{i\over 2}g\tau^a\Phi^i) +({\rm h.c.})
-g\varepsilon^{fac}W^c_{\nu}(W^{f[\nu,\mu]}
+g\varepsilon^{fde}W^d_{\mu}W^e_{\nu})\right] \cr} \eqno\eq
$$
and for the Higgs doublets
$$\eqalign{
\partial_{\mu}(D^{\mu}\Phi_1)^{\dagger}
-&(D^{\mu}\Phi_1)^{\dagger}\left(-{i\over 2}g{\bf \tau}.{\bf W}_{\mu}-{i\over
2}{\tilde g}'B_{\mu}\right) +2\lambda_1\Phi_1^{\dagger}(\Phi_1^{\dagger}\Phi_1
- {\nu_1^2 \over 2}) + \cr &2\lambda_3\Phi_1^{\dagger}[(\Phi_1^{\dagger}\Phi_1
- {\nu_1^2 \over 2}) + (\Phi_2^{\dagger}\Phi_2 - {\nu_2^2 \over 2})]
+\lambda_4[\Phi_1^{\dagger}(\Phi_2^{\dagger}\Phi_2)
-\Phi_2^{\dagger}(\Phi_1^{\dagger}\Phi_2)] + \cr &{\lambda_5\over
2}\Phi_2^{\dagger} [\Phi_1^{\dagger}\Phi_2 +\Phi_2^{\dagger}\Phi_1
-\nu_1\nu_2\cos\xi] + \cr &i{\lambda_6\over 2}\Phi_2^{\dagger}
[-i(\Phi_1^{\dagger}\Phi_2 -\Phi_2^{\dagger}\Phi_1) -\nu_1\nu_2\sin\xi] =0 \cr}
\eqno\eq
$$
and
$$\eqalign{
\partial_{\mu}(D^{\mu}\Phi_2)^{\dagger}
-&(D^{\mu}\Phi_2)^{\dagger}\left(-{i\over 2}g{\bf \tau}.{\bf W}_{\mu}-{i\over
2}{\tilde g}'B_{\mu}\right) +2\lambda_1\Phi_2^{\dagger}(\Phi_2^{\dagger}\Phi_2
- {\nu_2^2 \over 2}) + \cr &2\lambda_3\Phi_2^{\dagger}[(\Phi_1^{\dagger}\Phi_1
- {\nu_1^2 \over 2}) + (\Phi_2^{\dagger}\Phi_2 - {\nu_2^2 \over 2})]
+\lambda_4[\Phi_2^{\dagger}(\Phi_1^{\dagger}\Phi_1)
-\Phi_1^{\dagger}(\Phi_2^{\dagger}\Phi_1)]+ \cr &{\lambda_5\over
2}\Phi_1^{\dagger} [\Phi_1^{\dagger}\Phi_2 +\Phi_2^{\dagger}\Phi_1
-\nu_1\nu_2\cos\xi] - \cr &i{\lambda_6\over 2}\Phi_1^{\dagger}
[-i(\Phi_1^{\dagger}\Phi_2 -\Phi_2^{\dagger}\Phi_1) -\nu_1\nu_2\sin\xi] =0 \cr}
\eqno\eq
$$
Now, as we have mentioned before, the two-Higgs potential we are using contains
CP violation essential for electroweak baryogenesis scenarios. This may be made
explicit by writing the Higgs doublets as
$$
\Phi_1=(0, \phi_1)^T \ \ \ \ , \ \ \ \ \Phi_2=(0, \phi_2 e^{i\theta})^T
\eqno\eq
$$
where $\phi_1$, $\phi_2$, $\theta$ are real and $\theta$ is a CP-odd phase.

Since $\chi$ is a condensate we set $\chi=0$ outside the string core and, since
the Higgs fields do not provide sources for the first two components of the
$SU(2)$ fields we may consistently set $W^1=W^2=0$. With these considerations
we may make combinations of the equations of motion for $B_{\mu}$ and
$W^3_{\mu}$ to give
$$
\partial_{\nu}[(gB^{\nu}+g'W^{3\nu})^{,\mu} - (gB^{\mu}+g'W^{3\mu})^{,\nu}] =0
$$
outside the core, and
$$
\partial_{\nu}[(gW^{3\nu}-g'B^{\nu})^{,\mu} - (gW^{3\mu}-g'B^{\mu})^{,\nu}]
-2G(D^{\mu}\Phi^i)^{\dagger}(iG^2)\Phi^i =0
$$
where we have defined $G=\sqrt{g^2+g'^2}$.

Now, making the usual definitions of the photon field, $A^{\mu}$, and the
Z-boson via $GA=gB+g'W^3$ and $GZ=gW^3-g'B$ and expanding the covariant
derivatives, the above equations become
$$
\partial_{\nu}A^{[\nu,\mu]}=0 \eqno\eq
$$
outside the core, so that electromagnetism is free there, and
$$
\partial_{\nu}Z^{[\nu,\mu]} -2G^2(\phi_1^2 +\phi_2^2)Z^{\mu}
-4G^2\phi_2^2\partial^{\mu}\theta =0 \eqno\eq
$$
We also have the equations of motion for the Higgs components $\phi_1$
$$\eqalign{
\partial_{\mu}\partial^{\mu}\phi_1& -iGZ^{\mu}\partial_{\mu}\phi_1 -{1\over
4}G^2Z_{\mu}Z^{\mu}\phi_1 +2(\lambda_1+\lambda_3)\phi_1(\phi_1^2-{\nu_1^2\over
2}) +2\lambda_3\phi_1(\phi_2^2-{\nu_2^2\over 2}) + \cr
&\lambda_5\phi_2(\phi_1\phi_2\cos\theta -{\nu_1\nu_2\over
2}\cos\xi)e^{-i\theta} +i\lambda_6\phi_2(\phi_1\phi_2\sin\theta
-{\nu_1\nu_2\over 2}\sin\xi)e^{-i\theta} =0 \cr} \eqno\eq
$$
and $\phi_2$
$$\eqalign{
\partial_{\mu}\partial^{\mu}\phi_2& -iGZ^{\mu}\partial_{\mu}\phi_2 -{1\over
4}G^2Z_{\mu}Z^{\mu}\phi_2 -2i\partial_{\mu}\theta(\partial^{\mu}\phi_2 -{i\over
2}GZ^{\mu}\phi_2) -\phi_2\partial_{\mu}\partial^{\mu}\theta + \cr
&2(\lambda_2+\lambda_3)\phi_2(\phi_2^2-{\nu_2^2\over 2})
+2\lambda_3\phi_2(\phi_1^2-{\nu_1^2\over 2})
+\lambda_5\phi_1(\phi_1\phi_2\cos\theta -{\nu_1\nu_2\over 2}\cos\xi)e^{i\theta}
- \cr &i\lambda_6\phi_1(\phi_1\phi_2\sin\theta -{\nu_1\nu_2\over
2}\sin\xi)e^{i\theta} =0 \cr} \eqno\eq
$$

{\bf \chapter{Field Profiles Around the String}}
Given the general equations of motion derived in the previous section we may
now, consistent with the string ansatz, make the assumption of radial symmetry
and time-independence via $\phi_i=\phi_i(r)$, $Z^3=Z^3(r)$ and
$\theta=\theta(r)$. Our Z-equation becomes
$$
{1\over r}{d\over dr}\left(r{dZ^3\over dr}\right) -2G^2(\phi_1^2 +\phi_2^2)Z^3
=0 \eqno\eq
$$
Guided by the original work of Witten[16] and following Perkins and Davis[1],
we note that without the couplings to the fields outside the core $Z^3$ would
have a logarithmic profile with a scale governed by the current, I, along the
string. Since the couplings to the Higgs fields ``switch on" at some radius,
$r_0$, we shall use a thin-wall approximation for the Higgs fields. We shall
check the validity of this approximation numerically later.

We shall write
$$
\phi_i(r) = {1\over 2}\left(1+\tanh k_i(r-r_0)\right){\nu_i \over \sqrt{2}}
$$
Performing a Taylor expansion around $r=r_0$ of the Higgs fields in (3.1) we
obtain
$$
{1\over r}{d\over dr}\left(r{dZ^3\over dr}\right) = {1\over 4}G^2\nu^2Z^3
$$
which is the modified Bessel equation for $Z^3$. Thus, Z behaves like a
modified Bessel function in this region.

Using all the above information and demanding that $Z\rightarrow 0$ as
$r\rightarrow \infty$ we may make the following ansatz for Z
$$
Z^3(r)=\left\{\eqalign{ &\alpha \log\left({r\over \beta}\right)\ \ \ \ r<r_0
\cr &a K_0(G\nu r/2)\ \ \ \ r>r_0 \cr}\right.\eqno\eq
$$
where we have defined $2\nu^2 \equiv \nu_1^2 +\nu_2^2$ and $K_0$ is a modified
Bessel function. The parameter $\alpha$ is related to the current along the
string by $\alpha \equiv I/2\pi$ and is constant since we keep the current
fixed throughout this calculation.

Before proceeding further we shall specialize to the case where the vacuum
expectation values of the two doublets are equal, $\nu_1=\nu_2=\nu$. Our
thin-wall approximation is then
$$
\phi_1(r) = \phi_2(r) \equiv \phi(r) ={1\over 2}\left(1+\tanh
k(r-r_0)\right){\nu \over \sqrt{2}} \eqno\eq
$$
Note that this choice lies well within the accepted experimental bounds on the
two-Higgs model[15].

Our procedure is now as follows. We shall first impose continuity of $Z^3$ and
$dZ^3/dr$ at $r=r_0$ to reduce the number of arbitrary constants in (3.2) to
two, namely k and $r_0$, representing respectively the gradient of the Higgs
fields and the radius of electroweak symmetry restoration. We shall then
substitute these ansatze into the energy density and integrate to obtain the
total energy of the configuration. Finally we shall minimize the energy with
respect to our free parameters to obtain an estimate for $r_0$. The integration
of the relevant quantities is quite a formidable task. Fortunately for us it
will transpire that the extra terms we have compared to the analysis of the
one-Higgs case can be separated into two types. The first involves derivatives
of the CP-odd phase $\theta$ and is zero by virtue of the equations of motion
for the Higgs fields. The second type can be cast in a form identical to the
terms obtained for the one-Higgs calculation and thus acts only to modify the
parameters in that

 analysis. Thus we are able to perform corrections to the previous calculation
to obtain our result without having to explicitly carry out the integrations.

As in [1], continuity of $Z^3$ and $dZ^3/dr$ yields
$$
\log\left({\beta\over r_0}\right)={2\over \nu G r_0} \left({K_0(G\nu r/2) \over
 K_1(G\nu r/2)}\right) \eqno\eq
$$
where $K_0$ and $K_1$ are modified Bessel functions. Using this relation the
only remaining free parameters in the problem are k and $r_0$.

Now, the relevant terms in the energy density are
$$\eqalign{
\phi^2_{,r} +&{1\over 2}(Z^3_{,r})^2 +\phi^2\theta^2_{,r} +{1\over
2}G^2Z^2_3\phi^2
+(\lambda_1+\lambda_2+4\lambda_3)\left(\phi^4-\nu^2\phi^2+{1\over
4}\nu^4\right) \cr  &+\lambda_5\left(\phi^2\cos\theta-{1\over
2}\nu^2\cos\xi\right)^2
+\lambda_6\left(\phi^2\sin\theta-{1\over 2}\nu^2\sin\xi\right)^2 \cr}\eqno\eq
$$
Note that (2.10) implies that $\phi_2^2\theta_{,r}=0$ so we may drop this term,
as mentioned earlier. For the other terms we shall approximate $\theta$ by a
step-function
$$
\theta(r)=\left\{\eqalign{ &0 \ \ \ \ r<r_0 \cr & \xi\ \ \ \ r>r_0 \cr}\right.
\eqno\eq
$$
Since the Higgs fields are essentially zero for $r<r_0$ the final two terms in
(3.5) are of the same form as the other term in the Higgs potential part of the
energy density. Thus, we may just add these terms together so that each term is
now of the same form as those in the single Higgs case of Ref.~1. We may then
write
$$
E=E^{\Lambda}_{1H} \eqno\eq
$$
where $E^{\Lambda}_{1H}$ is defined to be the energy for the one-Higgs case
with the Higgs self-coupling, $\lambda$, replaced via
$$
\lambda\rightarrow\Lambda\equiv(\lambda_1+\lambda_2+4\lambda_3
+\lambda_5\cos^2\xi +\lambda_6\sin^2\xi) \eqno\eq
$$
At this stage we may just quote the modified result from Ref.~1. For large
current we look for a solution to the minimization equations with $r_0 \propto
\alpha$. The optimization equation for $r_0$ is
$$
{\partial E \over \partial r_0} = 0
$$
which, to leading order in $r_0$, yields
$$
{-16\Lambda\over 64}\nu^4r_0 +{\alpha^2\over 2r_0} =0 \eqno\eq
$$
where we have used the complementary equation, $\partial E/\partial k =0$, to
simplify the expression. Thus, defining $M_H=2\lambda\nu^2$ and $M_Z=G\nu/2$ we
have
$$
r_0 = \sqrt{{\lambda\over\Lambda}}{\alpha G\over M_H M_Z} \eqno\eq
$$
This is our final result. Compared to the result obtained in Ref.~1 for the
standard model we have a suppression prefactor of
$$
\left(\lambda \over
\lambda_1+\lambda_2+4\lambda_4+\lambda_5\cos^2\xi+\lambda_6\sin^2\xi\right)^{1/2}
$$
where $\lambda$ is the Higgs self-coupling in the standard model. For coupling
constants of order unity this is a suppression of less than an order of
magnitude. A diagram of the internal structure of the string, with the
electroweak symmetry restored out to a radius much larger than the core radius,
 is shown in figure~1. Since this calculation involves a series of
approximations we shall now numerically check the results in the standard model
in order to justify using the same approximations in this more complicated
setting.

{\bf \chapter{Numerical Approach in the Standard Model}}
Our final result, (3.10), is, of course, only meaningful if the various
approximations used to represent the field profiles and integrate the
Lagrangian density are valid. In particular there are three questions which we
focus on concerning the analytic approach.

First, the background ansatz (3.2) for the behaviour of the Z-field is based
partly on the fact that the string is superconducting and partly on the
assumption that the Higgs field behaves approximately linearly in the
neighbourhood of the restoration radius, $r_0$. In particular, given this
ansatz, it is not clear that the approximate form (3.3) will be valid for the
Higgs field. This is one assumption which we wish to check.

Secondly, in integrating the Lagrangian density there are several
approximations made along the way which we would like to support with a
numerical evaluation of the energy for various values of the restoration
radius. This should enable us to find the value of $r_0$ which minimizes the
energy of the string configuration.

Finally, in solving the optimization equations which arise from varying the
Lagrangian with respect to the free parameters of the theory we make the
assumption that $r_0 \propto I$, where $I$ is the superconducting current on
the string. This enables us to find analytic solutions to the equations.
Ideally we would like to be able to verify this assumption numerically. In
particular, by evaluating $r_0$ numerically for fixed current and then carrying
out this procedure for various values of the current it should be possible to
plot $r_0$ vs $I$ and thus to confirm or reject the assumptions underlying the
analytic calculation.

We shall address these questions in the context of the standard electroweak
model with one Higgs doublet and compare our results with those obtained
analytically in Ref.~1. In this way we hope to validate the assumptions
underlying our calculation in the two-doublet model since there we have applied
the same assumptions. The advantage of using the one-doublet model is clear
since the parameter space is much reduced and therefore significantly less work
is required.

Figure~2 shows a plot of the Higgs field (solid line) for the ansatz (3.2) for
the Z-field. The radius of symmetry restoration was set to be 45 (in arbitrary
units) and the Higgs equation of motion was solved using a relaxation method
subject to the boundary conditions $\phi(0)=0$ and $\phi(100)=1$. The dotted
line is a hyperbolic tangent as in the ansatz (3.3) with $r_0=45$ and $k=0.4$.
The agreement between the two curves is seen to be very good. A similar
analysis was carried out for widely varying values of $r_0$ and qualitatively
the same results were obtained in each case. Thus we conclude that the
thin-wall approximation (3.3) is a valid ansatz to make in this analysis.

Figure~3 shows a plot of the radius of symmetry restoration against the energy
of the configuration for fixed current along the string. This is obtained from
numerical integration of the energy density given in (3.5). The minimum of the
curve occurs at the optimal value of restored symmetry which we have referred
to as $r_0$. This analysis was carried out for a range of values of current and
the resulting plot of current, $I$, against restored symmetry radius, $r_0$, is
shown in figure~4. It is therefore verified that the relationship is linear for
large current and the assumptions underlying our analytic approach are valid.

{\bf \chapter{Conclusions}}
We have investigated the restoration of the electroweak symmetry around a
superconducting cosmic string in the two-Higgs doublet model. We have shown
that the more complex nature of the Higgs sector does not significantly affect
the results of Perkins and Davis[1]. The electroweak symmetry is restored out
to a radius comparable with the single Higgs case with the only difference
being a slight suppression due to factors of coupling constants and CP
violating phases. In particular, for sufficiently high currents the symmetry is
restored out to a macroscopic radius, which for GUT strings is of the order of
$10^{-5}$m.

Further, we have used a sequence of numerical methods to confirm the analytical
results of Ref.~1. We assume that the validity of the approximations used there
remains when the techniques are applied to the two-doublet model.

In the region of restored symmetry we may expect baryon violating effects to be
unsuppressed and the CP violation explicitly included in our calculation to
drive these processes preferentially in the direction of increasing baryon
number during the evolution of the string network. A description of such a
scenario for defect-mediated electroweak baryogenesis can be found in Ref.~5.
which provided the motivation for this study.

\medskip

\noindent{\bf Acknowledgements}

I am grateful to Robert Brandenberger and Anne Davis for useful discussions and
encouragement. I would also like to thank Andrew Sornborger and Stephen Hahn
for useful comments concerning the numerics. This work was supported in part by
the US Department of Energy under Grant DE-FG02-91ER40688, Task A.

\medskip

\noindent{\bf References}

\medskip
\pointbegin
W.B. Perkins and A-C. Davis, {\it Nucl. Phys.} {\bf B406}, 377 (1993).
\point
J. Ambj\o rn, N.K. Nielsen and P. Olesen, {\it Nucl. Phys.} {\bf B310}, 625
(1988);
J. Ambj\o rn and P. Olesen, {\it Intern. J. Mod. Phys.} {\bf A5}, 4525 (1990).
\point
P. Damgaard and D. Espriu, {\it Phys. Lett.} {\bf B256}, 442 (1991).
\point
V. Kuzmin, V. Rubakov and M. Shaposhnikov, {\it Phys. Lett.} {\bf B155}, 36
(1985); P. Arnold and L. McLerran, {\it Phys. Rev.} {\bf D36}, 581 (1987).
\point
R. Brandenberger, A-C. Davis and M. Trodden, BROWN-HET-935, DAMTP-94-6,
hep-ph\\9403215 (1994).
\point
A. Vilenkin, {\it Phys. Rev. Lett.} {\bf 46}, 1169 (1981).
\point
R. Brandenberger, L. Perivolaropoulos and A. Stebbins, {\it Int. J. Mod. Phys.}
{\bf A}, Vol. 5, No. 9, 1633 (1990).
\point
T. Vachaspati and A. Vilenkin, {\it Phys. Rev. Lett.} {\bf 67}, 1057 (1991).
\point
D. Vollick, {Phys. Rev.} {\bf D45}, 1884 (1992).
\point
N. Turok and T. Zadrozny, {\it Phys. Rev. Lett.} {\bf 65}, 2331 (1990); N.
Turok and J. Zadrozny, {\it Nucl. Phys.} {\bf B358}, 471 (1991); L. McLerran,
M. Shaposhnikov, N. Turok and M. Voloshin, {\it Phys. Lett.} {\bf B256}, 451
(1991).
\point
A. Cohen, D. Kaplan and A. Nelson, {\it Phys. Lett.} {\bf B263}, 86 (1991).
\point
A. Nelson, D. Kaplan and A. Cohen, {\it Nucl. Phys.} {\bf B373}, 453 (1992).
\point
M. Dine, R. Leigh, P. Huet, A. Linde and D. Linde, {\it Phys. Rev.} {\bf D46},
550 (1992).
\point
R. Brandenberger and A. Davis, {\it Phys. Lett.} {\bf B308}, 79 (1993).
\point
J.F. Gunion, H.E. Haber, G.L. Kane and S. Dawson, {\it The Higgs Hunter's
Guide}, (Addison-Wesley, Reading, MA 1989).
\point
E. Witten, {\it Nucl. Phys.} {\bf B249}, 557 (1985).
\point
H. Nielsen and P. Olesen, {\it Nucl. Phys.} {\bf B61}, 45 (1973).

\eject

\noindent{\bf Figure Captions}
\medskip
{\bf Figure 1} Diagram showing the string, its core, and the extent to which
the electroweak symmetry is restored. A profile for the electroweak Higgs field
is also shown.
\medskip
{\bf Figure 2} Solid line: Numerical solution for the Higgs field, $\phi(r)$,
around a superconducting cosmic string in the standard electroweak model with
the Z-field given by equation (3.2). Dotted line: a hyperbolic tangent as in
(3.3) with $r_0=45$ and $k=0.4$. Note how accurately these functions agree with
each other. Units are arbitrary.
\medskip
{\bf Figure 3} Total energy in the field configuration as a function of the
radius out to which the electroweak symmetry is restored, for fixed current
along the string. The minimum occurs at the value $r_0$. Units are arbitrary.
\medskip
{\bf Figure 4} Superconducting current along the string plotted against the
corresponding restored symmetry radius, $r_0$. The linear relationship for
large current is crucial to the analytic arguments in this paper. Units are
arbitrary.

\end